\documentclass[reprint,aps,prb,floatfix]{revtex4-2}
\usepackage{bm}
\usepackage{amssymb}
\usepackage{amsmath}
\usepackage{graphicx}
\usepackage{rotating}
\usepackage{epsfig}
\usepackage{psfrag}
\usepackage{amsmath}
\usepackage{subfigure}
\usepackage{braket}
\usepackage{tikz}
\usepackage{stmaryrd}
\usepackage{wasysym}
\usepackage{dsfont}

\newcommand{\bea}{\begin{eqnarray}}
\newcommand{\eea}{\end{eqnarray}}
\newcommand{\bpm}{\begin{pmatrix}}
\newcommand{\epm}{\end{pmatrix}}

\usepackage[unicode=true,colorlinks=true,pdfpagemode=UseOutlines,pdfstartview=Fith]{hyperref}
\hypersetup{linkcolor=blue,citecolor=blue,urlcolor=blue}

\begin{document}
\title{Emergent phases in the Yao-Lee model via coupling to topological spin textures}

\author{Muhammad Akram$^{1,2}$, Onur Erten$^1$}
\affiliation{$^1$Department of Physics, Arizona State University, Tempe, AZ 85287, USA \\ $^2$Department of Physics, Balochistan University of Information Technology, Engineering and Management Sciences (BUITEMS), Quetta 87300, Pakistan}

\begin{abstract}
Electrons in metals experience an effective vector potential when coupled to spin textures with non-zero scalar spin chirality, such as skyrmions. This coupling can generate a substantial field, leading to pronounced observable phenomena, including the topological Hall effect. Motivated by this, we consider a bilayer model in which the Majorana fermions in the Yao-Lee model on one layer interact with topological spin textures on the second layer via a spin-spin interaction. Unlike the Kitaev model, the Yao-Lee model remains exactly solvable, allowing us to perform Monte Carlo simulations to determine its ground state. Our analysis indicates that skyrmion crystals can give rise to a variety of vison crystals that are periodic arrangements of the $\mathbb{Z}_2$ fluxes with unusual patterns such as a kagome pattern. In addition, Majorana fermions acquire a substantial Berry phase from skyrmion crystals, resulting in phases with finite Chern numbers up to $\nu =5$. In the case of a single skyrmion defect in the magnetic layer, a corresponding defect in the vison configuration can be realized. These defects support localized states when the spin liquid is gapped. Similar to skyrmion crystals, spiral spin textures also give rise to a diverse range of flux crystals. However, in this case, most of these phases are gapless, with only a few being trivially-gapped. Our results highlight the rich physics emerging from the interplay between topological spin textures and fractionalized quasiparticles in quantum spin liquids.
\end{abstract}

\maketitle
\section{Introduction}
Skyrmions are topological defects of 2D magnets and have gained significant interest for their potential applications, such as racetrack memory devices\cite{Rossler2006}. In addition to their particle-like properties, skyrmions give rise to unique transport phenomena, including the topological Hall and Nerst effects\cite{PhysRevB.92.115417, PhysRevB.88.064409}. These phenomena arise from coupling conduction electrons to spin textures with finite scalar spin chirality, which generates an effective magnetic field. 

Motivated by these phenomena, we explore if similar effects can arise among fractionalized quasiparticles coupled to non-trivial spin textures. Quantum spin liquids (QSLs) provide an ideal platform for such exploration, as they lack magnetic ordering even at zero temperature while hosting fractionalized excitations and emergent gauge fields arising from long-range quantum entanglement.~\cite{ Balents_Nature2010, Zhou_RMP2017, Wen_RMP2017, Knolle_AnnRevCondMatPhys2019, Broholm_Science2020}. A key example is the Kitaev model on a honeycomb lattice~\cite{Kitaev_AnnPhys2006}, one of the first exactly solvable models with a QSL ground state. However, despite its theoretical elegance, the Kitaev model is highly sensitive to perturbations\cite{Rau_PRL2014}, and it is impossible to maintain exact solvability when coupled to an external field.

One way to address these issues is to extend the Kitaev model to include additional orbital degrees of freedom, leading to spin-orbital models\cite{Yao_PRL2011, Wu_PRB2009, Yao_PRL2009, Carvalho_PRB2018, Chulliparambil_PRB2020, Seifert_PRL2020,   Vijayvargia_PRR2023, Akram_PRB2023, Keskiner_PRB2023, Majumder_PRB2024,  Vijayvargia_arXiv2025, Keskiner_arXiv2025, Fontana_arXiv2025, Chulliparambil_PRB2021, Nica_npjQM2023, Poliakov_PRB2024, Wu_PRL2024, Mandal_APL2024}, such as the Yao-Lee (YL) model\cite{Yao_PRL2011}. The YL model features an enlarged local Hilbert space, which provides greater flexibility in incorporating perturbations while preserving the model's exact solvability. Previous studies on the YL model have examined the effects of interlayer coupling, Dzyaloshinskii-Moriya interaction, external magnetic fields and non-Hermitian effects, revealing a rich variety of distinct phases\cite{Nica_npjQM2023,Chulliparambil_PRB2021,Poliakov_PRB2024, Wu_PRL2024, Mandal_APL2024}.

\begin{figure}[t]
\center
\includegraphics[width=\columnwidth]{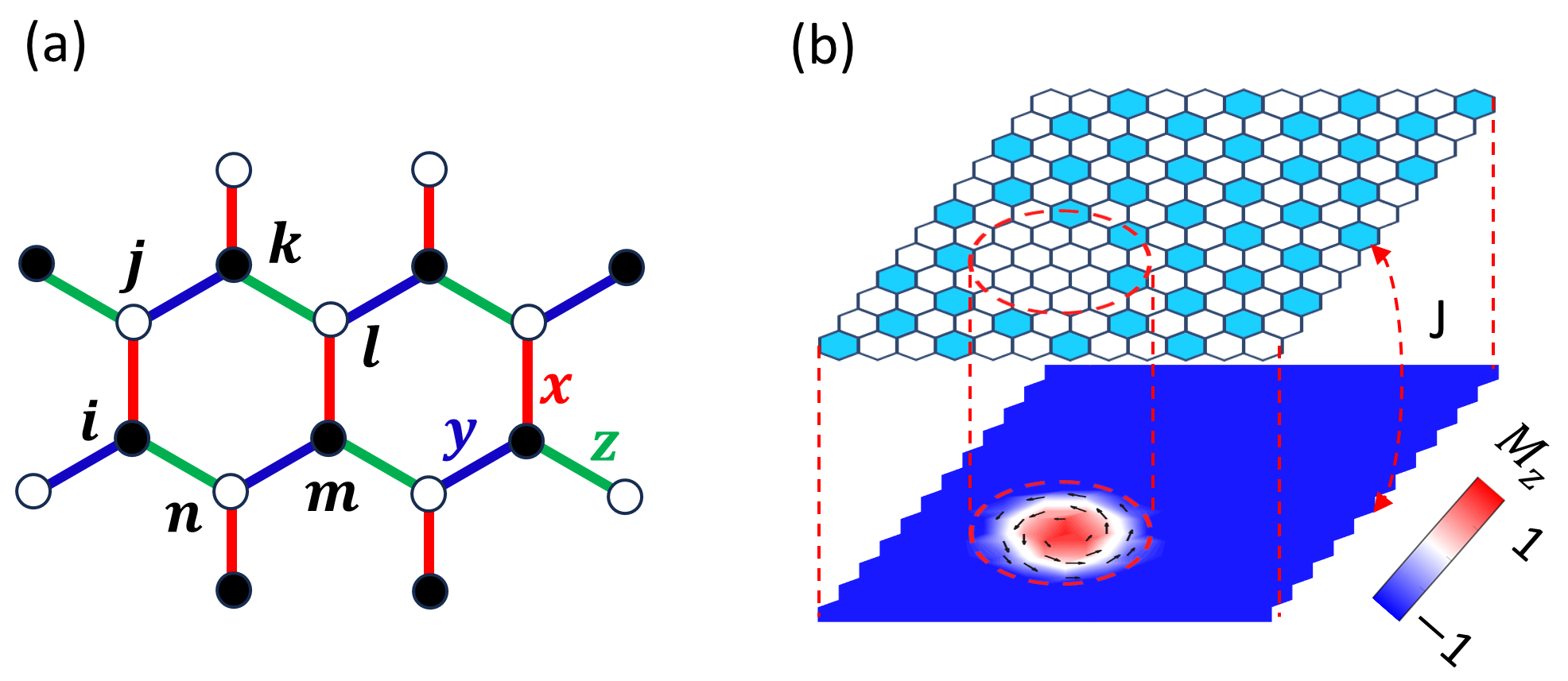} 
\caption{Schematic of the Yao-Lee model on a honeycomb lattice, showing three types of bonds x, y, and z represented by red, blue, and green colors, respectively. (b) Schematic of the Yao-Lee model coupled to an adjacent layer with a spatially varying magnetization with a single skyrmion defect. The color code on the top layer shows the 0 and $\pi$ fluxes (white and blue, respectively), and in the bottom layer, it represents the z-component of the magnetization, and the arrows represent the in-plane magnetization components.}
\label{Fig:1}
\end{figure}

This work investigates a bilayer model consisting of a YL model on one layer coupled to a magnetic layer exhibiting non-colinear spin textures as depicted in Fig.~\ref{Fig:1}(a, b). The YL model retains its exact solvability even in the presence of non-collinear spin textures and we perform Monte Carlo simulations and variational analysis to determine its ground state. Our key findings are as follows: (i) The non-colinear spin textures mix different flavors of Majorana fermions in the YL model and can lead to the formation of $\mathbb{Z}_2$ fluxes. (ii) For a skyrmion crystal, the phase diagram exhibits a diverse range of vison crystals, which are the periodic arrangements of $\mathbb{Z}_2$ fluxes as a function of skyrmion size and the interlayer coupling. Most of these vison crystals are gapped and topological, with Chern numbers varying from 0 to 5. (iii) A single skyrmion defect in the magnetization pattern gives rise to a corresponding defect in the vison configuration of the spin liquid ground state. (iv) Similarly, for a spiral configuration, the phase diagram also displays a variety of vison crystals. However, most of these vison crystals are gapless in this case.

The rest of the article is organized as follows. In Section II, we introduce the model and describe the Majorana fermion representation. In Section III, we present our results for (a) skyrmion crystal pattern, (b) a single skyrmion defect and (c) spiral magnetization. We conclude with a summary of our results and an outlook.  

\begin{figure*}[t]
\center
\includegraphics[width=\textwidth]{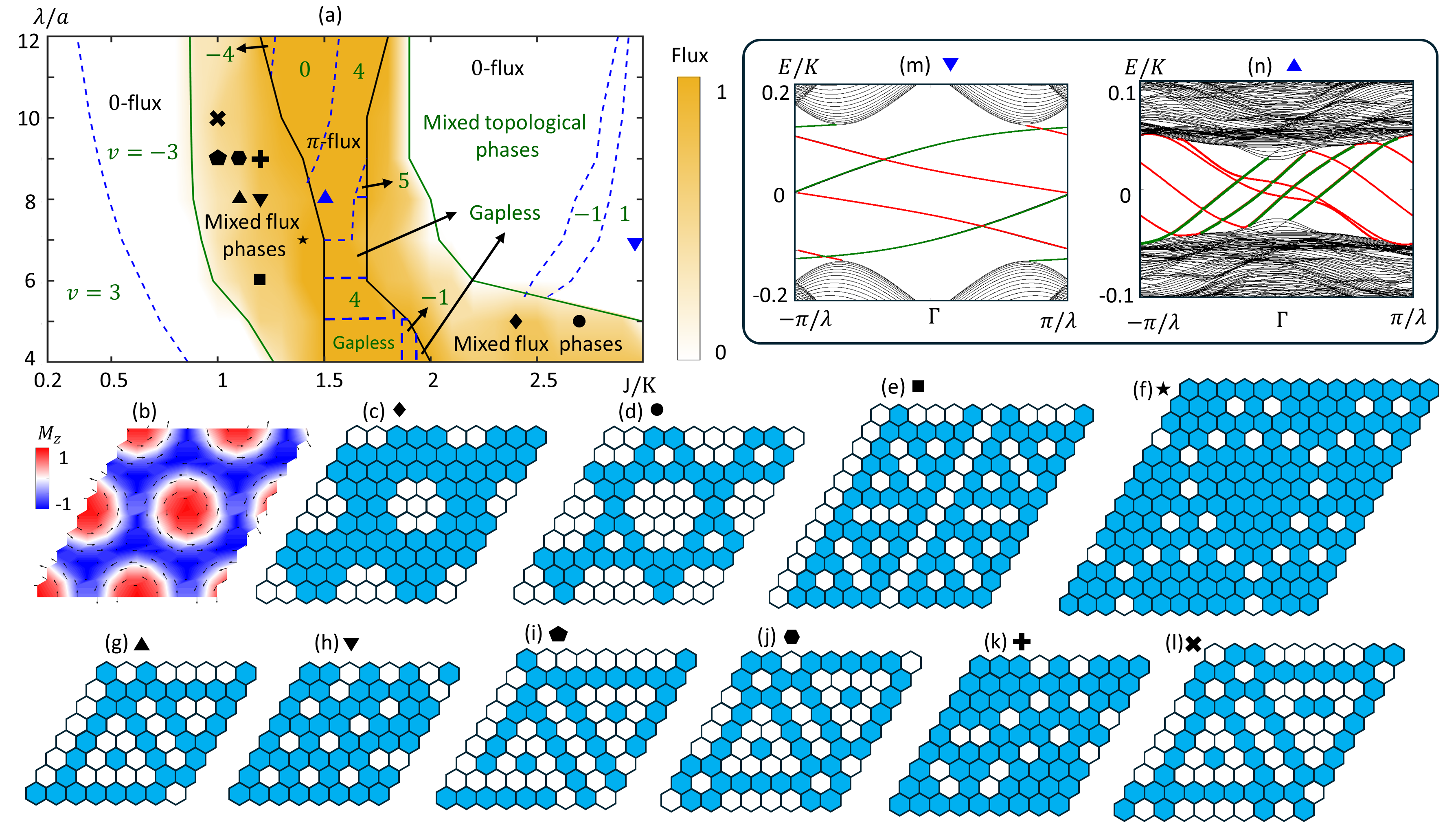} 
\caption{(a) Phase diagram of vison crystals for skyrmion crystal magnetization texture as a function of the interlayer coupling $J$ and wavelength of skyrmions $\lambda$. The solid green lines indicate the phase transitions between the 0-flux and mixed phases, while the solid black lines represent the transitions between the $\pi$-flux and mixed phases. The dashed blue lines denote the topological phase transitions. (b) Illustration of a $2\times2$ skyrmion crystal system with a skyrmion size of $\lambda/a=5$. The color represents the
z-component of magnetization, and the arrows represent the in-plane magnetization components. (c-f) Vison crystals for $2\times2$ skyrmion crystal systems: (c) $J/K =2.4$, $\lambda/a=5$; (d) $J/K =2.7$, $\lambda/a=5$; (e) $J/K =1.2$, $\lambda/a=6$; (f) $J/K =1.4$, $\lambda/a=7$. (g-l) Flux crystals for $1\times1$ skyrmion crystal systems: (g) $J/K =1.1$, $\lambda/a=8$; (h) $J/K =1.2$, $\lambda/a=8$; (i) $J/K =1$, $\lambda/a=9$; (j) $J/K =1.1$, $\lambda/a=9$; (k) $J/K =1.2$, $\lambda/a=9$; (l) $J/K =1$, $\lambda/a=10$. (m) Chiral edge states in the 0-flux, $\nu=1$ phase at $J/K =3$ and $\lambda/a=7$. The red and green colors represent edge states on the top and bottom open boundaries respectively. (n) Chiral edge modes for the $\pi$-flux, $\nu=4$ phase at $J/K=1.5$ and $\lambda/a=8$.}
\label{Fig:2}
\end{figure*}

\section{Model and Methods}
We consider a bilayer honeycomb model as shown in Fig.~\ref{Fig:1}(a, b). The first layer is described by the YL model\cite{Yao_PRL2011}, while the second has a classically ordered magnetic state. The two layers are ferromagnetically coupled via a spin-spin interaction (such as Hund's coupling). We assume that the interlayer coupling is weak compared to the intralayer interactions on the second layer and the classically ordered state is unaffected by the interlayer interaction. As such, the second layer acts as a source for a non-uniform field on the first layer. The Hamiltonian for the YL layer is $H= H_{YL}+H_{J}$ where
\begin{eqnarray}
H_{YL}&=& \sum_{\langle ij \rangle_\alpha} K^{\alpha} \left(\tau_{i}^{\alpha} \tau_{j}^{\alpha}\right) 
\left( \pmb{\sigma}_{i} \cdot \pmb{\sigma}_{j} \right), \\
H_{J}&=& -J\sum_i \pmb{M}_{i} \cdot \pmb{\sigma}_{i}.
\end{eqnarray}

Here, $K^{\alpha}$ represents the nearest-neighbor coupling constant for the $\alpha$ bond, $\pmb{M}_{i}$ is the magnetization of the second layer at site $i$, and $|\pmb{M}_{i}|^2=1$. A key aspect of the YL model is the presence of conserved Wilson loop operators, $W$ defined around each plaquette as shown in Fig.~\ref{Fig:1}(a). These plaquette operators are defined as $W= \tau_i^x\tau_j^y\tau_k^z\tau_l^x\tau_m^y\tau_n^z\otimes \mathds{1}$. Note that since $W$ only involves only orbital degrees of freedom (DOFs) and it commutes with both $H_{YL}$ and $H_J$ and therefore also with the total Hamiltonian, $[H,~W]=0$. Consequently, the eigenvalues of the plaquette operator, $W=\pm 1$, can be used to classify the eigenstates of the Hamiltonian.

It is possible to obtain an exact solution to the Hamiltonian by introducing six Majorana fermions per site for the spin and orbital DOF\cite{Yao_PRL2011}: $\sigma_{j}^{\alpha}=-i\epsilon^{\alpha \beta \gamma}c_{j}^{\beta}c_{j}^{\gamma}/2$, $\tau_{j}^{\alpha}=-i\epsilon^{\alpha \beta \gamma}d_{j}^{\beta}d_{j}^{\gamma}/2$ and $\sigma_i^\alpha \tau_j^\beta=ic_i^\alpha d_j^\beta$. This representation is redundant, and the physical states are the eigenstates of operators $D_{i}=-i c_{i}^{x} c_{i}^{y} c_{i}^{z} d_{i}^{x} d_{i}^{y} d_{i}^{z}$ with eigenvalues $1$. The constraints can be enforced via the projection operators: $
P = \prod_i (1 + D_{i})/2.$
In the Majorana representation, the Hamiltonian is expressed as $H = P \mathcal{H} P$, where \(\mathcal{H}=\mathcal{H}_{YL}+\mathcal{H}_{J}\),
\begin{eqnarray}
\mathcal{H}_{YL}&=&\sum_{\langle ij \rangle_\alpha} K^{\alpha} u^{\alpha}_{ij}[ic_{i}^{x} c_{j}^{x}+i c_{i}^{y} c_{j}^{y} + ic_{i}^{z} c_{j}^{z}] \\
\mathcal{H}_{J}&=&\sum_{i} iJ[ M_{i}^xc_{i}^{y} c_{i}^{z}+i M_{i}^yc_{i}^{z} c_{i}^{x}+i M_{i}^zc_{i}^{x} c_{i}^{y}].
\end{eqnarray}
Here $u^{\alpha}_{ij}=-id_{i}^{\alpha}d_{j}^{\alpha}$ is the bond operator and commutes with $\mathcal{H}$. In addition, the plaquette operators can be expressed as products of the bond operators $W=u_{ij}^xu_{jk}^yu_{kl}^zu_{lm}^xu_{mn}^yu_{ni}^z$. 

According to Lieb's theorem\cite{Lieb_PRL1994}, the ground state of the YL model lies in the 0-flux sector. In addition, the excitation spectra of the three flavors of Majorana fermions are identical, which can be obtained via Fourier transformation over half of the Brillouin zone. The spectra are gapped when the $K^\alpha$ along one bond exceeds the sum of the interactions along the other two bonds; otherwise, they are gapless. In our analysis, we consider isotropic Kitaev interactions with $K_x=K_y=K_z=K$. Under these conditions, the Majorana fermions exhibit Dirac-like behavior and are coupled to static $\mathds{Z}_2$ gauge fields. Flipping a bond operator in the 0-flux state induces two $\pi$-flux excitations, commonly referred as visons. Visons affect the kinetic energy of Majorana fermions by introducing a $\pi$ Berry phase when the Majorana fermions move around a $\pi$-flux plaquette. A periodic arrangement of such visons is called a vison crystal. 

In the presence of an effective magnetic field, Lieb's theorem is violated, and 0-flux is not necessarily the ground state. Ref.~\citenum{Chulliparambil_PRB2021} explored the phase diagram of the YL model with a uniform field and showed that the ground state shifts from 0 to 1/3, $\pi$ and back to 0-flux with increasing field.

YL model requires a four-dimensional local Hilbert space and has bond-directional interactions. Recently, Ref.~\citenum{Churchill_npjQM2025} proposed a microscopic route to realize YL model based on Kugel-Khomskii type superexchange mechanism for partially filled $e_g$ orbitals. 

To obtain the ground state flux configuration, we perform classical Monte Carlo simulations with 15,000 Monte Carlo steps per temperature down to $T=10^{-3}K$. We repeat the simulations 40 times for mixed-flux phases to ensure the ground state is stable.  
We also applied a variational analysis to the flux crystals obtained from the Monte Carlo simulations and constructed the phase diagram for the spiral case. 

\section{Results and Discussion}

\subsection{Skyrmion crystal magnetization texture}
We first consider a skyrmion crystal magnetization as depicted in Fig.~\ref{Fig:2}(b). The skyrmion crystals are formed by a superposition of three spiral states that are rotated by $2\pi/3$ with respect to each other. The wavelength of the spiral, $\lambda$, corresponds to the diameter of the skyrmions. Fig.~\ref{Fig:2}(a) shows the phase diagram as a function of the interlayer coupling $J/K$ and $\lambda$. For small $J$, the ground state stays in the 0-flux sector. As $J$ increases, the average flux continuously increases and various mixed phases with different average fluxes emerge. The average flux increases until it reaches a $\pi$-flux phase, and the $\pi$-flux phase remains stable over a certain range of $J$. As $J$ increases further, the average flux decreases, leading to various mixed phases, and finally returns to 0 flux.  The range of mixed phases is more pronounced for small skyrmions and decreases for larger skyrmions. The shape of the flux configurations varies with changes in both $J$ and $\lambda$. When $\lambda$ is commensurate with certain patterns, flux crystals with well-defined shapes emerge, some of which are shown in Fig.~\ref{Fig:2}(c-l). Unusual flux patterns such as a `flower' pattern (Fig.~\ref{Fig:2}(c)) and a kagome pattern (Fig.~\ref{Fig:2}(d)) can be stabilized. In Appendix~\ref{AppendixVC}, Fig.~\ref{Fig:1:supp}, we present several more vison crystal configurations with larger system sizes.

\begin{figure*}[t]
\center
\includegraphics[width=\textwidth]{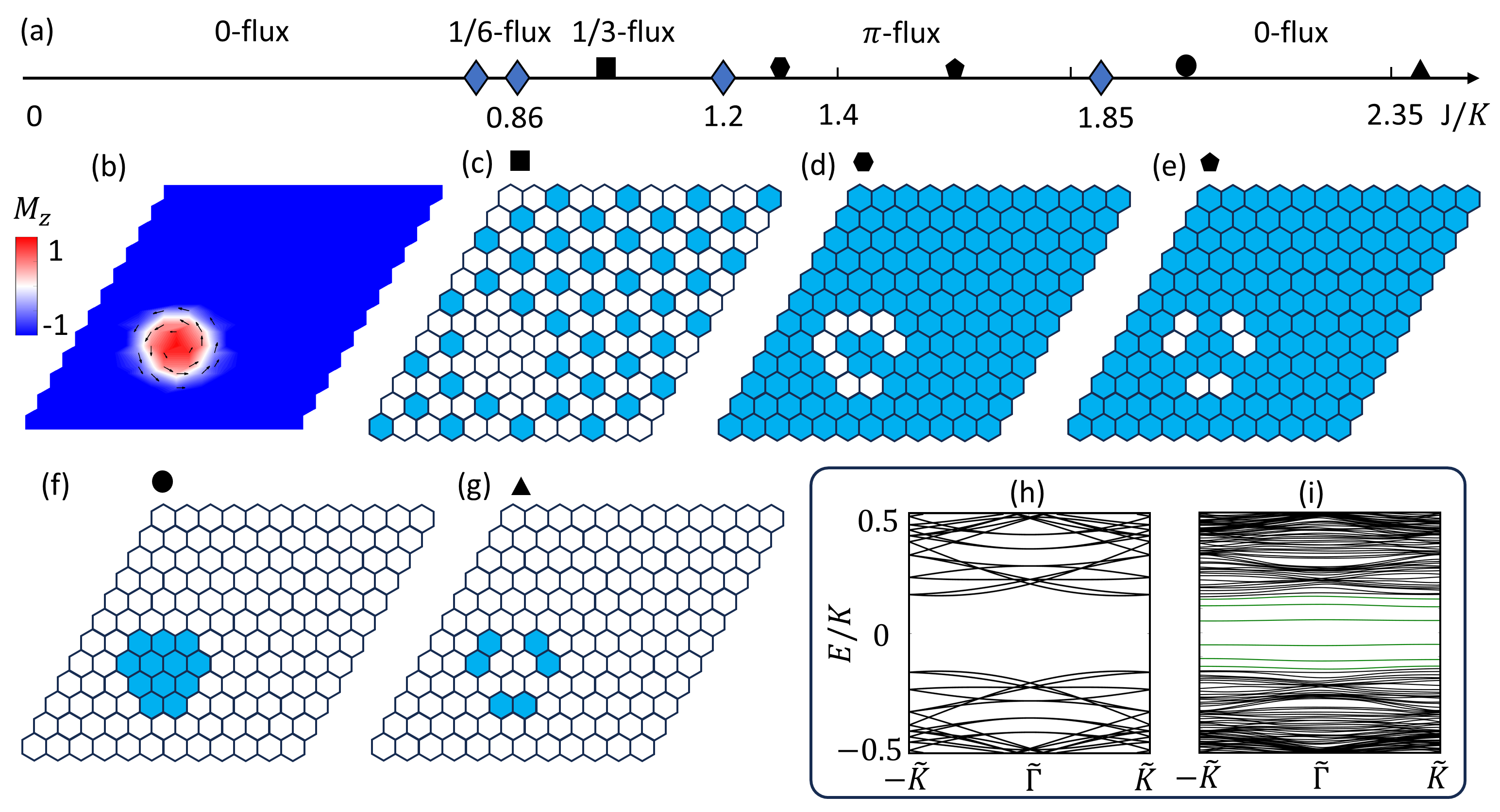} 
\caption{(a) Phase diagram of the YL model coupled to the FM magnetization with a skyrmion defect as a function of the interlayer coupling $J$. (b) Illustration of the FM magnetization with a skyrmion defect. The system size is $12a\times12a$, and the skyrmion diameter is $\lambda/a=4$. (c-g) Fluxes with defects on top of the skyrmions in FM magnetization; (c) $1/3-$flux at $J/K =1$; (d) $\pi-$flux at $J/K =1.2$; (e) $\pi-$flux at $J/K =1.4$; (f) 0-flux at $J/K =2$; (g) 0-flux at $J/K =2.4$. Majorana fermion excitation spectrum of $1/3-$flux crystal at $J/K =1$ (h) without and (i) with a skyrmion defect. Here $\tilde{K}$ and $\tilde{\Gamma}$ denote the corner and center of the reduced Brillouin zone, respectively.  Green lines in (i) represent the localized states at the skyrmion defect.}
\label{Fig:3}
\end{figure*}

The dominant phases in the phase diagram are the 0-flux and $\pi$-flux phases. Both phases exhibit gapped and gapless regimes, with the gapped regime further dividing into trivial and topological phases. The topological phases are stabilized by the Berry phase imparted by the skyrmions on the Majorana fermions. Similar effects were predicted for triangular Kondo lattice magnets with chiral magnetic order\cite{Martin_PRL2008}. These topological phases are distinguished by their Chern number
\begin{eqnarray}
\nu=\frac{1}{\pi}\sum_{\mu,\nu} \int_{BZ/2} d^2k~ \text{tr}F_{xy}^{\mu,\nu}(k)
\end{eqnarray}
of the Bogoliubov-de Gennes Hamiltonian of the Majorana fermions. Here $F_{xy}^{\mu,\nu}=\partial_{k_x} A_y^{\mu,\nu}-\partial_{k_y}A_x^{\mu,\nu} +i([A_x,A_y])^{\mu,\nu}$ is the Berry curvature, $A^{\mu,\nu}=-i\langle \pmb{n}^\mu(k)|\nabla_k|\pmb{n}^\nu(k)\rangle$ is the non-Abelian Berry connection and $\mu$,$\nu$ represent filled bands defined over half Brillouin zone\cite{murakami20042,Chulliparambil_PRB2021}. For small $J$, the 0-flux phase exhibits two topological phases with $\nu=3$ and -3. On the other hand, for large $J$, the significant topological phases are $\nu=1$ and -1, which emerge large $J$. In addition, other topological phases appear in narrow regions, which we describe as mixed topological phases. Similar to the 0-flux, the $\pi$-flux also exhibits both gapped and gapless spectra. In the gapped phase, we observe topological phases with Chern numbers $\nu=0,1,-1, -4,4,5$. 

\subsection{Ferromagnetic magnetization with a single skyrmion defect}
Next, we consider a uniform ferromagnetic (FM) magnetization with an isolated single skyrmion defect. Our phase diagram obtained from Monte Carlo simulations in the absence of skyrmion defects agree well with previous variational analysis\cite{Chulliparambil_PRB2021} except we discover a small region of 1/6-flux as shown in Fig.~\ref{Fig:3}(a). Next, we include the skyrmion defect and explore its effects. We consider a $12a\times12a$ system, while the skyrmion has a diameter of $\lambda/a =4$, as depicted in Fig.~\ref{Fig:3}(b). 

At small interlayer coupling, within the 0-flux sector, the presence of the defect has negligible impact. The flux configuration around the skyrmion remains unchanged and the Fermi surface is largely unaffected. As $J$ increases, the flux distribution over the skyrmion deviates from the flux over the uniform magnetization, leading to a defect. For 1/6 and 1/3-fluxes, the resulting defects are 0-flux, as illustrated in Fig.~\ref{Fig:3}(c) and Fig.~\ref{Fig:2:supp} in Appendix~\ref{AppendixVB}. For the $\pi-$flux, the resulting defects are formed over the skyrmion's boundary, and the skyrmion's center remains $\pi-$flux. For $1.2<J/K<1.4$, the defect consists of a string of 0-fluxes, as shown in Fig.~\ref{Fig:3}(d). However, for $J/K>1.4$, three pairs of 0-fluxes appear along the boundary, related by $C_3$ symmetry, as shown in Fig.~\ref{Fig:3}(e). For $J/K>1.4$, there are no defects in pi-flux. As $J$ increases further, the system exhibits 0 flux accompanied by the appearance of two types of defects. For 
$J/K<2.35$, there is a $\pi$-flux defect formed on the skyrmion, as illustrated in Fig.~\ref{Fig:3}(f). However, for $J/K>2.35$, three pairs of visons appear on the boundary of the skyrmion, related by $C_3$ symmetry, as shown in Fig.~\ref{Fig:3}(g).

The spectra of Majorana fermions coupled to a FM magnetization remain gapless for 0, $\pi$, and 1/6-fluxes. However, the spectrum is gapped for the 1/3-flux. Fig.~\ref{Fig:3}(i) and Fig.~\ref{Fig:3}(h) illustrate the band structures for FM magnetization with and without a skyrmion defect respectively. In the gapped phase, additional flat bands emerge within the energy gap, as shown in Fig.~\ref{Fig:3}(i). These flat bands correspond to localized states around the 0-flux defect and arise due to the presence of the skyrmion. Note that the localized states are gapped, similar to the vortex bound states in trivial superconductors\cite{Tinkham_book}.

\begin{figure}[t]
\center
\includegraphics[width=\columnwidth]{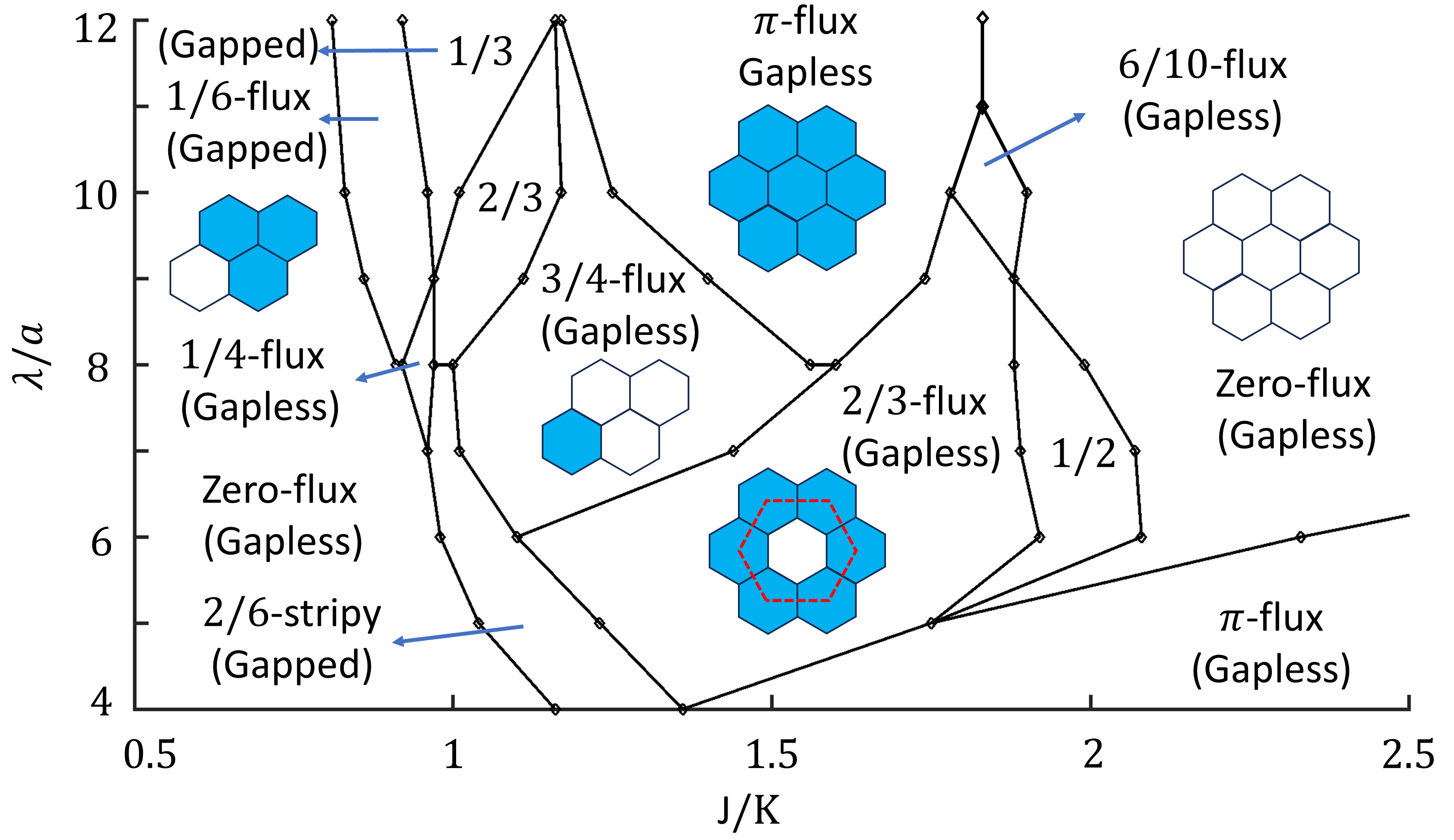} 
\caption{Phase diagram of flux crystals as a function of the interlayer coupling $J$ and wavelength of spiral $\lambda$ for spiral magnetization texture.}
\label{Fig:4}
\end{figure}

\subsection{Spiral magnetization texture}
Next, we consider a coplanar spiral magnetization characterized by a single wave vector. Although coplanar spin textures do not generate an emergent magnetic field for metals, they have been shown to produce nontrivial Berry phase effects in superconductors\cite{Nadj_PRB2013}. Our Monte Carlo simulations suggest that 11 different flux crystals can be stabilized, as shown in Fig.~\ref{Fig:3:supp}, Appendix~\ref{AppendixV}. In order to accurately determine the phase boundaries among these phases, we calculate their energies variationally and construct the phase diagram as shown in Fig.~\ref{Fig:4}.
For large wavelengths ($\lambda/a\sim12$), our results match with the flux configurations observed under a uniform magnetic field\cite{Chulliparambil_PRB2021}, as the effective magnetic field varies slowly in space. In this regime, we find 0, $\pi$, 1/3, and 1/6 flux crystals. However, as the wavelength decreases, the results deviate from those of the uniform magnetic field. The dominant phases in the phase diagram are the 0 and $\pi$-fluxes, though 2/3 and 3/4-fluxes also occupy significant regions of the phase diagram. Additionally, 
1/2, 1/5, 2/6-stripy, and 6/10-fluxes are also stabilized in certain regions of the phase diagram, as illustrated in the Fig.~\ref{Fig:4}. The excitation spectra of the Majorana fermions are gapless for most of the vison crystals. However, a few vison crystals are gapped with $\nu=0$, including the 1/3, 1/6, and 2/6-stripy fluxes. 

\section{Conclusion}
We studied the phase diagram of the YL model coupled to noncollinear spin textures, including skyrmion crystals, isolated skyrmion defects, and spiral magnetization. YL model stays exactly solvable in the presence of these spin textures which allows us to perform classical Monte Carlo simulations to determine the ground state vison configurations. We showed that the skyrmion crystals can give rise to unconventional flux patterns and gapped phases that are characterized by a tunable Chern number reaching up to $\nu=5$. FM magnetization with single skyrmion defects introduces corresponding defects in the flux configuration and can trap localized states. Spirals can also give rise to a variety of flux configurations, yet most are gapless and the few gapped phases have $\nu=0$. Interesting future directions include coupling QSLs to time-dependent spin textures and Floquet engineering dynamically generated Chern bands.

\section{Acknowledgments}
We acknowledge support from NSF
Award No. DMR-2234352. We thank the ASU Research Computing Center for high performance computing resources.

%To get further insight, consider a coplanar spiral in the yz-plane with the q-vector along the x-axis. The magnetization varies along the x-axis and we rotate the quantization axis from the fixed z-axis to the axis parallel to the magnetization via local transformation:
%\[U = e^{-i \frac{\theta}{2} \hat{\sigma} \cdot \hat{n}} = \cos\left( \frac{\theta}{2} \right) I - i \sin\left( \frac{\theta}{2} \right) \hat{\sigma} \cdot \hat{n} \]
%where $\hat{n} = \frac{\hat{z} \times \hat{M}}{|\hat{z} \times \hat{M}|}=-\hat{x}$ and $\theta = \cos^{-1} \left( \hat{M} \cdot \hat{z} \right)$. Under these tranformations the Majorana fermions transform as 
%\[c_x \rightarrow c_x \\
%c_y \rightarrow \cos(\theta) c_y - \sin(\theta) c_z\\
%c_z \rightarrow \cos(\theta) c_z + \sin(\theta) c_y\].
%Using these transformations, the first term of Kitaev interactions %$ic_xc_x$ is unchanged, whereas the second and third terms of the %Kitaev interactions transform as
%\[c_i^y c_{i+1}^y + c_i^z c_{i+1}^z \rightarrow \cos(\Delta \theta) \left( c_i^y c_{i+1}^y \\ + c_i^z c_{i+1}^z \right) - \sin(\Delta \theta) \left( c_i^y c_{i+1}^z - c_i^z c_{i+1}^y \right) \]. Thus, in the new basis with the quantization axis aligned with the magnetization, the Kitaev interaction transforms into a reduced Kitaev interaction proportional to $\cos(\Delta \theta)$ and an emergent Dzyaloshinskii-Moriya interaction (DMI) proportional to $\sin(\Delta \theta)$ along the q vector of the spiral.

%\newpage
%\clearpage
\appendix
\section{Vison crystals for skyrmion crystal magnetization texture}\label{AppendixVC}
\begin{figure*}
\center
\includegraphics[width=\textwidth]{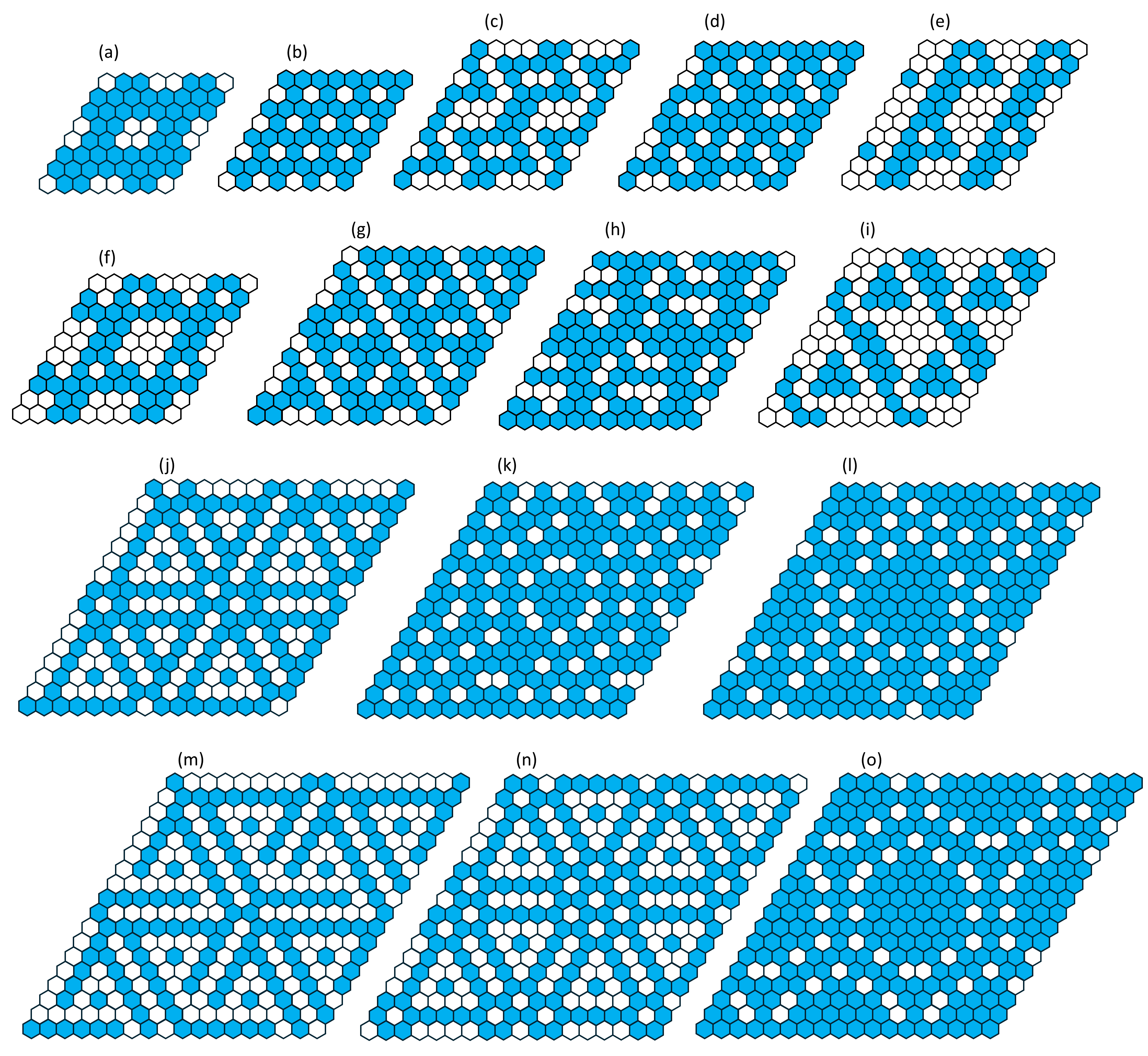} 
\caption{Vison crystals obtained from skyrmion crystal magnetization texture for $2\times2$ skyrmion systems: (a) $J/K =2.1$, $\lambda/a=4$; (b) $J/K =2.5$, $\lambda/a=4$; (c) $J/K =1.3$, $\lambda/a=5$; (d) $J/K =1.4$, $\lambda/a=5$; (e) $J/K =2.2$, $\lambda/a=5$; (f) $J/K =2.3$, $\lambda/a=5$; (g) $J/K =1.3$, $\lambda/a=6$; (h) $J/K =1.9$, $\lambda/a=6$; (i) $J/K =2$, $\lambda/a=6$; (j) $J/K =1.1$, $\lambda/a=8$; (k) $J/K =1.2$, $\lambda/a=8$; (l) $J/K =1.8$, $\lambda/a=8$; (m) $J/K =1$, $\lambda/a=9$ for $2\times2$; (n) $J/K =1.1$, $\lambda/a=9$; (o) $J/K =1.2$, $\lambda/a=9$.}
\label{Fig:1:supp}
\end{figure*}

In Fig.~\ref{Fig:1:supp}, we present some of the vison crystals that are obtained for skyrmion crystal magnetization texture. We use a $2\times 2$ unit cell which helps to visualize the periodicity and 3-fold rotation symmetry of the underlying skyrmion crystal. 

\section{1/6-flux crystal with 0-flux defect} \label{AppendixVB}

\begin{figure*}
\center
\includegraphics[width=0.6\textwidth]{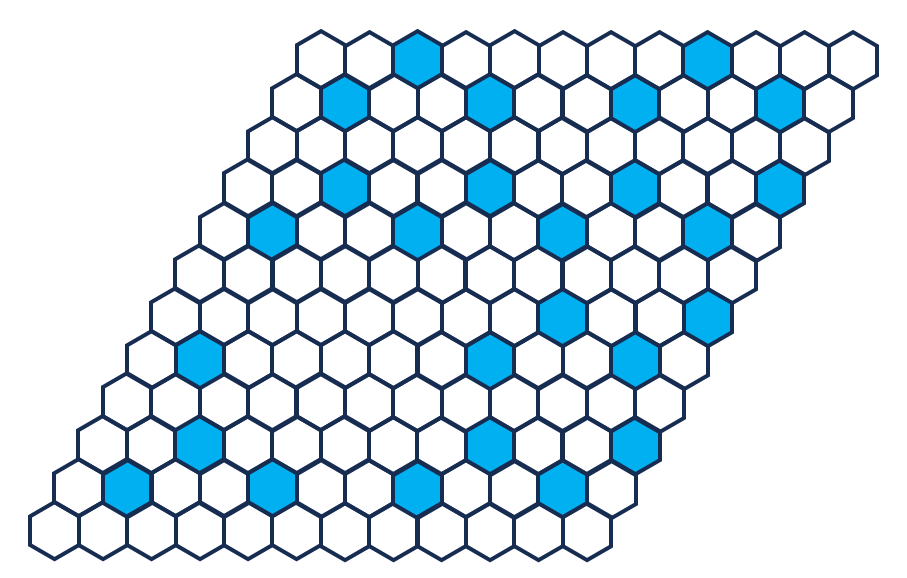} 
\caption{1/6-flux crystal with a 0-flux defect on top of the skyrmion in the FM magnetization.}
\label{Fig:2:supp}
\end{figure*}

Fig.~\ref{Fig:2:supp} shows 1/6-flux crystal with a 0-flux defect, which emerges in a ferromagnetic magnetization state with a skyrmion defect. While the uniform magnetization background generates the 1/6-flux configuration, the 0-flux defect is precisely localized at the skyrmion position. The system has dimensions of $12a\times12a$, containing a skyrmion of size $4a\times4a$.

\section{Variational Analysis} \label{AppendixV}

\begin{figure*}
\center
\includegraphics[width=\textwidth]{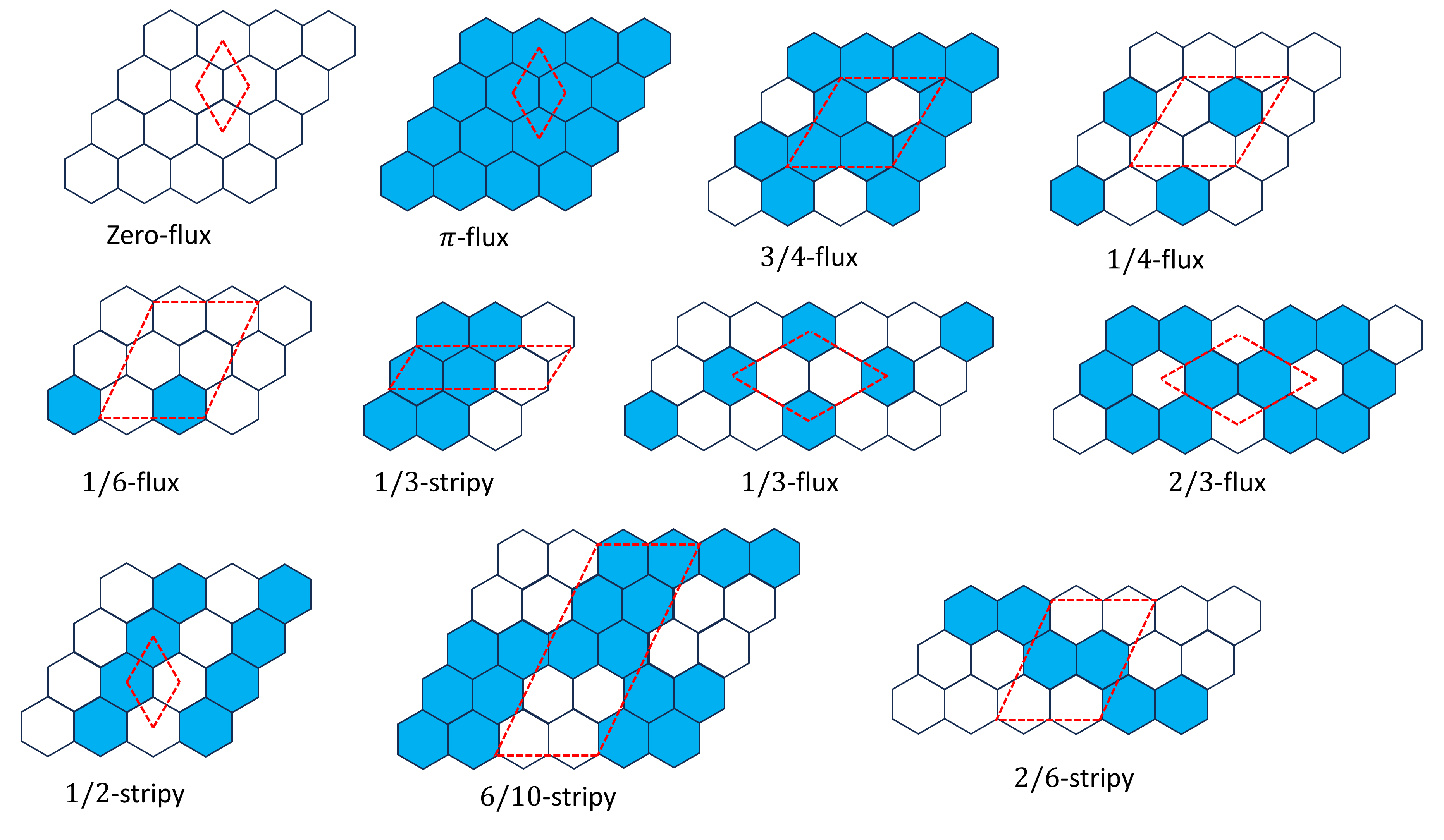} 
\caption{Vison crystals considered in the variational analysis for spiral magnetization. The white and sky-blue plaquettes correspond to 0 ($W=1$) and $\pi$ ($W=-1$) fluxes, respectively. The dashed lines denote the unit cell boundaries.}
\label{Fig:3:supp}
\end{figure*}

For the spiral background, we initially employed Monte Carlo simulations and identified eleven vison crystals, as illustrated in Fig.~\ref{Fig:3:supp}. Subsequently, we compared the energies of these configurations using variational analysis to construct the phase diagram. For each point on the phase diagram, we computed the energy of each vison crystal and compared these energies to construct the phase boundaries.

\bibliographystyle{apsrev4-1}
\bibliography{References}
\end{document}